\documentclass{article}
\usepackage{ijcai13}
\usepackage{epsfig}
\usepackage{amssymb}
\usepackage{amsmath}
\usepackage{amsfonts}
\usepackage[numbers]{natbib}

\usepackage{algorithm}
\usepackage{algorithmic}

\newtheorem{theorem}{Theorem}
\newtheorem{lemma}{Lemma}
\newtheorem{corollary}{Corollary}
\newtheorem{proposition}{Proposition}

\usepackage{times}

\title{Reinforcement Ranking}
\author{Hengshuai Yao and Dale Schuurmans \\
Department of Computer Science\\
University of Alberta\\
Edmonton, AB, Canada T6G2E8 \\
hengshua,dale@cs.ualberta.ca}

\begin{document}

\maketitle

\begin{abstract}
We introduce a new framework for web page ranking---reinforcement
ranking---that improves the stability and accuracy of Page Rank while
eliminating the need for computing the stationary distribution of random walks.  
Instead of relying on teleportation to ensure a well defined Markov chain, 
we develop a reverse-time reinforcement learning framework that determines
web page authority based on the solution of a reverse Bellman equation.  
In particular, for a given reward function and surfing policy we recover a
well defined authority score from a reverse-time perspective: looking
back from a web page, what is the total incoming discounted reward brought 
by the surfer from the page's predecessors?  
This results in a novel form of reverse-time 
dynamic-programming/reinforcement-learning problem that
achieves several advantages over Page Rank based methods:
First, stochasticity, ergodicity, and irreducibility of the underlying Markov chain is no longer required for well-posedness.
Second, the method is less sensitive to graph topology and more stable in the presence of dangling pages.
Third, not only does the reverse Bellman iteration yield a more efficient power 
iteration, it allows for faster updating in the presence of graph changes.
Finally, our experiments demonstrate improvements in ranking quality.

\end{abstract}

\section{Introduction}

{\em Page Rank} is a dominant link analysis algorithm for web page ranking
\cite{Page98,hits,liu07},
which has been applied to a wide range of problems in
information retrieval and social network analysis
\cite{moorePageRank,Adheat,IPMcitationpr,Backstrom:2011:SRW:1935826.1935914}.~Under Page Rank, authoritativeness 
is defined by the stationary distribution of a Markov chain 
constructed from the web link structure 
\cite{qdpr,jeh03,PR_survey,pr_survey3,revisit06,liu07,ordinalpr}. 
On each page, a model surfer follows a random link, jumping to the 
linked page and continuing to follow a random link.
Thus, pages are treated as arriving in a Markov chain---the next page 
visited depends only on the page where the surfer currently visits. 
The rank of a web page is then defined as the probability of visiting the 
page in a long run of this random walk. 
Unfortunately, this simple protocol does not allow the surfer to proceed from
a page that has no outgoing links---such pages are called \emph{dangling} pages.
In these cases, the Markov chain derived from the link structure of the Web is
not necessarily irreducible or aperiodic, which are required 
to guarantee the existence of the stationary distribution. 
To circumvent these problems, a {\em teleportation} operator is introduced 
that allows the surfer to escape dangling pages 
by following artificial links added to the Web graph.
Teleportation has been widely adopted by literature, 
leading to the well accepted stationary distribution formulation 
of authority ranking, see e.g.\ 
\cite{TaherAda,PR_extra,jeh03,pr_c1,qdpr,PR_inout}.

However, if we consider real search behavior, teleportation is obviously 
artificial.
It is unnatural to propagate the score of a page to other unlinked pages,
thus teleportation contributes a blind regularization effect rather than 
any real information.
In fact, teleportation contradicts the basic hypothesis of Page Rank:
through teleportation, pages that are not linked by a page still receive 
reinforcement from the page.
Teleportation was primarily introduced to guarantee 
the existence of the stationary distribution.
In this paper, we show that teleportation is in fact unnecessary for 
identifying authoritative pages on the Web.
First, contrary to widely accepted belief, teleportation is not required to
derive a convergent power iteration for global Page Rank style authority scores.
Second, as has been widely adopted in the random surfer interpretation for 
{\em Page Rank}, teleportation or even random walk is also unnecessary conceptually. 
We introduce a new approach to defining web page authority
that is based on a novel reinforcement learning model
that avoids the use of teleportation while remaining well defined.
We prove that the authority function is well posed and satisfies a 
reverse Bellman equation.
We also prove that the induced reverse Bellman iteration, 
which is more efficient than the Page Rank procedure, 
is guaranteed to converge for any positive discount factor.

In addition to establishing theoretical soundness, we also show that the 
reinforcement based authority function is less sensitive to link changes. 
This allows us to achieve faster updates under graph changes, addressing
the Page Rank {\em updating} problem \cite{Bahmani:2012:PEG:2339530.2339539} in an efficient new way.
As early as 2000, it was observed that $23\%$ of the web pages 
changed their index daily \cite{searchweb}. 
Unfortunately, the Page Rank power iteration does not benefit
significantly from initialization with the previous stationary
distribution \cite{amy06}. 
We prove that our authority function can take better advantage of
initialization, and yield faster updates to graph changes.
Furthermore, 
we demonstrate that reinforcement ranking can 
improve on 
the authority scores
produced by Page Rank in a controlled case study.

\section{Page Rank}

We first briefly review the formulation of Page Rank
\cite{PR_survey,pr_survey3,revisit06,liu07}.
Suppose there are $N$ pages in the Web graph under consideration.
Let $L$ denote the adjacency matrix of the graph;
i.e., $L(i,j)=1$ if there is a link from page $i$ to page $j$, 
otherwise $L(i,j)=0$.
Let $H$ denote the row normalized matrix of $L$;
let $e$ be the vector of all ones;
and let $v$ denote the {\em teleportation vector}, 
which is a normalized probability vector
(assume column vectors).
Finally, let $S$ be a stochastic matrix such that $S=H+a u^T$, 
where the vector $a$ indicates $a_i=1$ if page $i$ is dangling 
and $0$ otherwise.
Here $u$ is a probability vector that is normally set to either $e/N$ or $v$.
Note that adding $a  u^T$ to the $H$ matrix artificially 
``patches'' the dangling pages that block the random surfer. 

The transition probability matrix used by Page Rank is 
\[
G=c S+ (1-c) ev^T,
\]
for a convex combination parameter $c\in(0,1)$.
The matrix $G$ is stochastic, irreducible and aperiodic, 
and thus its stationary distribution exists and is unique according 
to classical Markov chain theory. 
In fact, the Page Rank (denoted by $\bar{\pi}$) is precisely the stationary 
distribution vector for $G$, which satisfies $\bar{\pi}=G^T{\bar{\pi}}$.
Page Rank can be interpreted as follows:
with probability $c$ the surfer follows a link,
otherwise with probability $1-c$ the surfer teleports to a page 
according to the distribution $v$;
the rank of a page is then given by its long run visit frequency.
Teleportation is key, since it ensures the chain is irreducible 
and aperiodic, thus guaranteeing 
the existence of a stationary distribution for the surfing process. 

Unfortunately, the introduction of 
teleportation causes the matrix $G$ to become completely dense. 
Power iteration is therefore impractical unless one exploits 
its special structure in $G$; namely that 
it is a sparse plus two rank one matrices.
An efficient procedure for computing Page Rank is given
in Algorithm \ref{alg:pipi} \cite{TaherAda,PR_extra}.
This algorithm evaluates an equivalent update to 
$\bar{\pi}_{k+1} =G^{T}\bar{\pi}_{k}$,
but it avoids using $G$ by implicitly incorporating
the scores of the dangling pages and teleportation in computing $\omega$.
Note that the issue of accommodating dangling pages in Page Rank has 
been considered a challenging research issue 
\cite{pr_frontier,PR_survey}.

\begin{algorithm}[t]
   \caption{Standard procedure for computing Page Rank: efficient power iteration method that exploits $G$'s structure.
   }
   \label{alg:pipi}

    {Initialize $x_{0}$ }   

    {\bfseries repeat} 
   {

	\quad $x_{k+1} = cH^{T}x_{k}$
        
       \quad $\omega = ||x_{k}||_{1} -  ||x_{k+1}||_{1} $
        
	\quad $x_{k+1}= x_{k+1} + \omega v$

}

 {\bfseries until desired accuracy is reached} 
   
\end{algorithm}

\section{MDPs and the Value Function}
\label{sec:mdp}

A Markov Decison Process (MDP) is defined by a 5-tuple 
$(\mathbb{S},\mathbb{A},\mathcal{P}^{\mathbb{A}},\mathcal{R}^{\mathbb{A}},\gamma)$;
where $\mathbb{S}$ denotes a state space;
$\mathbb{A}$ is the action space; 
$\mathcal{P}^{\mathbb{A}}$ is a transition model 
with $\mathcal{P}^{a}(s,s')$ 
being the probability of transitioning to state $s'$ after taking
action $a$ at state $s$; 
$\mathcal{R}^{\mathbb{A}}$ is a reward model with $\mathcal{R}^{a}(s,s')$ 
being the reward of taking action $a$ in state $s$ and transitioning 
to state $s'$;
and $\gamma \in [0,1)$ is a discount factor
\cite{puterman94,rl_survey,Bert96ndp,sutton98book}.

A {\em policy} $\pi$ maps a state $s$ and an action $a$ into a probability 
$\pi(s,a)$ of choosing that action in the state. 
The \emph{value} of a state $s$ under a policy $\pi$ is the discounted 
long-term future rewards received following the policy
\[
V^{\pi}(s)=E_{\pi} \Big\{\textstyle\sum_{t=0}^{\infty} \gamma ^t r_{t} \Big| s_0=s, a_{t\ge 0}\sim \pi \Big\},
\]
where $r_{t}$ is the reward received by the agent at time $t$, 
and $E_{\pi}$ is the expectation taken with respect to the distribution 
of the states under the policy.

The value function satisfies an equality called {\em Bellman equation}. 
In particular, for any state $s \in \mathbb{S}$
\begin{equation}\label{bellman}
V^{\pi}(s) = \gamma \textstyle\sum_{s'\in \mathbb{S}} P^{\pi}(s,s')  V^{\pi}(s') + \bar{r}^{\pi}(s),
\end{equation}
where $P^{\pi}(s,s')$ is the probability of transitioning from 
$s$ to $s'$ following the policy, 
and the $\bar{r}^{\pi}(s)$ is the expected immediate reward of 
leaving state $s$ following the policy.



\section{The Reinforcement Ranking Framework}

We now introduce the {\em reinforcement ranking} framework, 
which models search and ranking in terms of an MDP. 
The framework is composed of the following elements. 

{\em The agent and the environment}.
The agent is a surfer model and the environment 
is a set of hyperlinked documents on which the surfer explores. 
That is, we consider the Web to be the environment;
the surfer acts by sending requests that are processed by servers on the Web.  
This is a simple model of everyday surfing
that stresses the subjectiveness of the surfer as well as the objective
structure of the Web, in contrast to Page Rank
which models surfing as a goal-less random walk.

{\em The rewards}.
According to \cite{sutton98book}, a reward function ``maps each perceived 
state (or state-action pair) of the environment to a single number, a reward, 
indicating the intrinsic desirability of that state''. 
Intuitively, a {\em reward} is a signal that evaluates an action.  
A surfer can click many hyperlinks on a page. 
If a clicking leads to a page that satisfies the surfer's needs, 
then a large reward is received; otherwise it incurs a small reward. 
From the perspective of information retrieval, the reward represents
information gain from reading a page. 
The introduction of rewards to search and ranking is important because 
it highlights the fact that a page has {\em intrinsic} importance to users.
In fact, this is a key difference from what has been pursued in the 
link analysis literature, which does not normally model pages as having 
intrinsic values.
The reward hypothesis is also important because surfing and search is 
purposeful in this model---{\em actions are taken to achieve rewards}. 
%
In this paper, we will be considering a special reward function, 
in which $\mathcal{R}^{a}(s,s') = r(s')$, 
where $r$ is a function mapping from the state space to real numbers. 
This means the reward of transitioning to a state is uniquely determined 
by the state itself. 

{\em The actions and the states}.
An {\em action} is the click of a hyperlink on a page. 
A {\em state} is a web page.
The current state is the current page 
being visited by the surfer.
After a clicking on a hyperlink, 
the surfer can observe the linked page or a failed connection.
For simplicity, we assume that all links are good in this paper.  
That is, the {\em next state} is always the page that an action leads to. 
Therefore, the state space $\mathbb{S}$ is the set of the web pages. 
The action space on a page $s$, denoted $\mathbb{A}(s)$, 
is the set of actions that lead to the linked pages from $s$. 
The overall action space is defined by the union of the actions available on
each page, i.e., $\mathbb{A} = \cup_{s\in \mathbb{S}}\mathbb{A}(s)$.

{\em The surfing policy and transition model}.
A surfer policy specifies how hyperlinks are followed at web pages.
Based on the above definitions relating web search to an MDP model,
we can equate a surfer with a standard
MDP policy as specified in Section \ref{sec:mdp}.
For web search, we also assume the transitions are {\em deterministic};
that is, clicking a hyperlink on a particular page always leads to the same 
successor page, hence
$\mathcal{P}^{a}(s,s')=1$ for all $a\in \mathbb{A}(s)$. 
This treatment simplifies the problem without losing generality---%
it is straightforward to extend our work to the other cases.

\subsection{The Authority Function}

Given these associations established between web surfing and an MDP, 
we can now develop a web page authority function in the framework of 
reinforcement ranking.
In particular,
we define the authority score of a page to be the rewards accumulated 
by its predecessors under the surfing policy.
That is, for a page $s\in \mathbb{S}$, its authority score under surfing 
policy $\pi$ is
\begin{eqnarray}
R^{\pi}(s) 
&=& 
r(s) +\gamma r^{(1)}(s)+ \gamma^2 r^{(2)}(s) + \cdots 
\nonumber 
\\
&=& 
\sum_{k=0}^{\infty} \gamma^k r^{(k)} (s), 
\label{R1}
\end{eqnarray}
%
where $\gamma$ is the discount factor, 
$r(s)$ is a reward that is dependent on $s$, 
and 
$ r^{(k)}(s)$ is the reward carried from the $k$-step predecessors of $s$ to 
$s$ by the policy.
Note that in the second equation, $r^{(0)} =r$,
and if a page $s$ has no predecessor, 
$R^{\pi}(s)$ can be set to $r(s)$ 
or some other default value.
The $k$-step historical rewards to a state $s$ are defined as follows:
\\
$r^{(1)}(s)$ captures the one-step rewards propagated into $s$
\[
r^{(1)}(s) = \sum_{p \in \mathbb{S}} P^{\pi}_{p,s} r(p)\,;
\]
$r^{(2)}(s)$ captures the rewards from the two-step predecessors
\[
r^{(2)}(s) = 
\sum_{p \in \mathbb{S}} P^{\pi}_{p,s}
\sum_{p' \in \mathbb{S}} P^{\pi}_{p',p} r(p')\,; 
\]
$r^{(3)}(s)$ captures the 3-step rewards propagated into $s$
\[
r^{(3)}(s) = 
\sum_{p \in \mathbb{S}} P^{\pi}_{p,s}
\sum_{p' \in \mathbb{S}} P^{\pi}_{p',p}   
\sum_{p'' \in \mathbb{S}} P^{\pi}_{p'',p'} r(p'');
\quad\mbox{etc.}
\]

Note that the discount factor $\gamma$ plays an important role in this model, 
since it controls the effective horizon over which reward is accumulated.
If $\gamma$ is large, 
the authority score will consider long chains of predecessors that lead
into a page.
If $\gamma$ is small, 
the authority score will only consider predecessors that are a within
a few steps of the page.
This gives a new interpretation for the dampening-factor-like in PageRank. 
Previously it is commonly recognized that the larger the dampening factor in PageRank, 
the closer the score vector reflects the true link structure of the graph, 
e.g. see \cite{pr_c_func,pr_c1,pr_cgeneric,pr_func2,pr_c}. 
While the two interpretations do not contradict each other,
viewing the dampening/discount factor as a control over the distance of looking back from pages
is surely both essential and intuitive.

\subsection{The Reverse Bellman Equation}

Although the authority score function $R^\pi$ appears to be similar 
to a standard value function $V^\pi$, they are not isomorphic concepts: 
the value function (\ref{bellman}) is defined in terms of the forward 
accumulated rewards.
The reverse function (\ref{R1}) cannot be reduced to the forward definition
(\ref{bellman}) because the transition probabilities are not normalized
in both directions; they are only normalized in the forward direction.
In particular, (\ref{bellman}) is an expectation, whereas (\ref{R1}) 
cannot be an expectation in general.
Despite this key technical difference, it is interesting (and ultimately
very useful) that the authority function also satisfies a reverse form of
Bellman equation.

\begin{theorem}[Reverse Bellman Equation]\label{thm:Rbell}
The authority function $R^\pi$ satisfies the reverse Bellman equation
for all $s$:
\begin{equation}\label{Rbell}
{R}^{\pi}(s) = \gamma \sum_{p\in \mathbb{S}}   P^{\pi}_{p,s} {R}^{\pi}(p) + r(s) .
\end{equation}
%
\end{theorem}

\noindent
{\bf Proof}:
First observe that the $k$-step rewards can be expressed in terms of the 
$(k-1)$-step rewards; that is
\begin{eqnarray*}
r^{(1)}(s) = \sum_{p\in \mathbb{S}} P^{\pi}_{p,s} r(p)\,,
\quad
r^{(2)}(s) = \sum_{p \in \mathbb{S}} P^{\pi}_{p,s} r^{(1)}(p)\,,
\quad
\mbox{etc.}
\end{eqnarray*}
Therefore, from the definition of $R^\pi$ in 
(\ref{R1}), 
one obtains
\begin{eqnarray}
{R}^{\pi}(s)
\!\!&\!\!=\!\!&\!\! 
r(s) + \gamma \left[r^{(1)}(s) +  \gamma r^{(2)}(s) + \ldots \right] 
\nonumber 
\\
\!\!&\!\!=\!\!&\!\! 
r(s) + \gamma \Big[\sum_{p\in \mathbb{S}} P^{\pi}_{p,s} r(p) 
+  \gamma \sum_{p \in \mathbb{S}} P_{p,s} r^{(1)}(p) + \ldots \Big] 
\nonumber 
\\
\!\!&\!\!=\!\!&\!\! 
r(s) + \gamma \sum_{p\in \mathbb{S}} P^{\pi}_{p,s} \left[ r(p) 
+  \gamma  r^{(1)}(p) + \ldots \right] 
\nonumber 
\\
\!\!&\!\!=\!\!&\!\!
r(s) + \gamma \sum_{p\in \mathbb{S}}   P^{\pi}_{p,s} {R}^{\pi}(p). 
\nonumber
\quad \square
\end{eqnarray}
%

The standard Bellman equation (\ref{bellman})
looks forward from a state to define its value, 
but equation (\ref{Rbell}) looks backward from a state in define its authority.
Therefore, we call equation (\ref{Rbell}) the 
{\em reverse Bellman equation} ({\em RBE} for short). 
Similar to Page Rank, ${R}^{\pi}$ determines the authority of a page
based on its back links.
However, ${R}^{\pi}$ is well defined without 
teleportation.
In particular, the surfer model $P^{\pi}$ is defined on the link structure only,
without any teleportation.
Notice that $P^{\pi}$ is not necessarily irreducible or aperiodic,
in fact it is not even stochastic on rows for dangling pages.
Yet, perhaps surprisingly, one is still able to achieve a well defined
authority score ${R}^{\pi}$, 
which is not possible from the classical Markov chain theory.

\begin{theorem}\label{thm:welldef}
For $\gamma \inÊ [0, 1)$, any policy $\pi$ and any bounded reward 
function $r$, $R^{\pi}$ is finite. 
\end{theorem}

\noindent
{\bf Proof}:
It can be shown that the spectral radius of $\gamma P^{\pi}$ is strictly 
smaller than that of any well defined policy. 
Therefore $I- \gamma (P^{\pi})^{T}$ is invertible. 
Additionally,
$(I- \gamma (P^{\pi})^{T})^{-1} = \sum_{t=0}^{\infty}(\gamma (P^{\pi})^{T})^t$
by \cite[Theorem 1.5]{saad03}.
Therefore, $R^{\pi} = (I- \gamma (P^{\pi})^{T})^{-1} r$, 
hence $R^{\pi}$ is finite for any policy and any bounded reward function $r$. 
$\square$


The practical significance of the RBE is that it yields an efficient 
algorithm for computing $R^\pi$, based on 
a backward version of value iteration
as used in dynamic programming and reinforcement learning;
see Algorithm \ref{alg:Rpi}.
To establish the correctness of this algorithm we first need a lemma.
Let $\|\cdot \|$ be the L-2 norm, 
$\|R_1\| = (\sum_{i=1}^N R_1(i)^2)^{1/2}$. 

\begin{lemma}\label{lem:contraction}
For any $R \in \mathbb{R}^N$, we have 
$\|(P^{\pi})^{T}R\| \le  \|R\|$. 
\end{lemma}

\noindent
{\bf Proof}:
~
\vspace*{-2\baselineskip}

\begin{eqnarray*}
\|(P^{\pi})^{T}R\|^2 
\!\!&\!\!=\!\!&\!\! 
\sum_{i=1}^N  \Big( \sum_{h=1}^N P^{\pi}_{hi} R(h)  \Big)^2
\le \sum_{i=1}^N  \sum_{h=1}^N P^{\pi}_{hi} R(h)^2
\\
\!\!&\!\!=\!\!&\!\! 
\sum_{h=1}^N  \sum_{i=1}^N  P^{\pi}_{hi} R(h)^2
= \sum_{h=1}^N  R(h)^2 =\|R\|^2.\square
\end{eqnarray*}

Note here we used the ordinary L-2 norm rather than the weighted L-2 norm, 
as is common in reinforcement learning. 

\begin{theorem}
For $\gamma\in[0,1)$ and finite $r$,
Algorithm \ref{alg:Rpi}'s update
has a unique fixed point to which the iteration must converge.
\end{theorem}

\noindent
{\bf Proof}:
The proof follows the Banach fixed-point theorem 
given in
\cite{Bert96ndp}. 
Define $T^{\pi}: \mathbb{R}^N \to \mathbb{R}^N$ be a mapping by, $ T^{\pi}(R^{\pi}) = \gamma (P^{\pi})^{T}R^{\pi} + r$.
$T^{\pi}$ is a contraction mapping in the L-2 norm because 
\begin{align*}
\|T^{\pi}(R_1)- T^{\pi}(R_2)\| &= \gamma\| P^{\pi} (R_1-R2)\|
\le 
\gamma \|R_1-R_2\|, 
\end{align*}
according to Lemma \ref{lem:contraction}.
It follows that the iteration converges to the unique
fixed point $R^{\pi} = T^{\pi}(R^{\pi})$. 
$\square$


This approach to computing an authority ranking 
has several advantages over Page Rank.
First, Algorithm \ref{alg:Rpi}
does not compute an additional $\omega$ factor
(which requires $2N$ additional flops per iteration). 
Second, no special treatment is required for dangling pages, 
which has generally been considered tricky for Page Rank
\cite{pr_frontier,PR_survey}. 
Finally, 
there is a significant improvement in computation cost
and sensitivity for Algorithm \ref{alg:Rpi} over Algorithm \ref{alg:pipi}.

\begin{algorithm}[t]
   \caption{Reverse Bellman iteration for computing $R^{\pi}$:
   no special treatment is required for dangling pages. 
   }
   \label{alg:Rpi}

    {Initialize $R_{0}$ }   

    {\bfseries repeat} 
   {
        
	\quad $R_{k+1}= \gamma ({P^{\pi}})^{T}R_{k} +  r$

}

 {\bfseries until desired accuracy is reached}
   
\end{algorithm}

\section{Sensitivity}

\begin{figure}[t]
\centering
\includegraphics[width=3.2in]{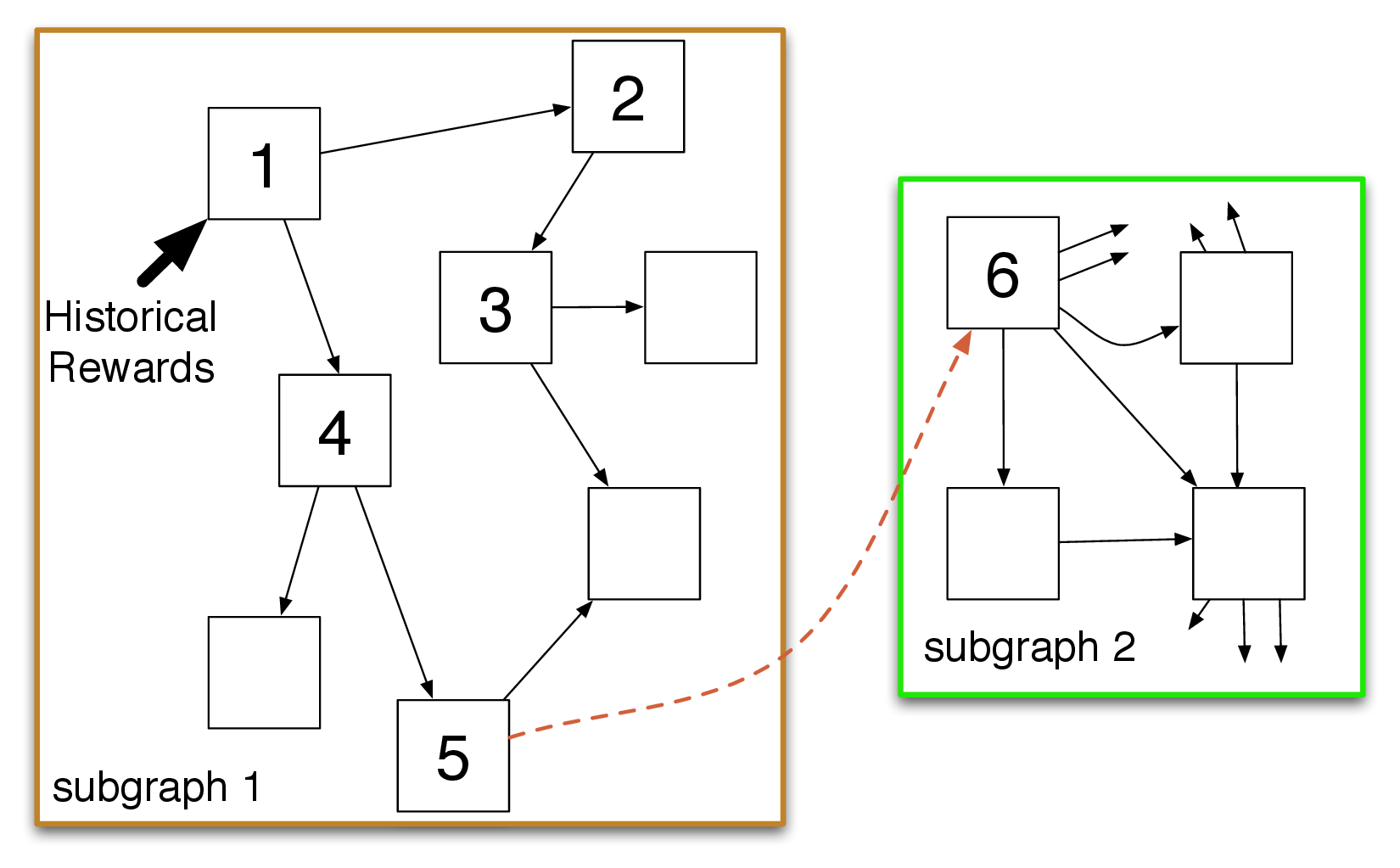}
\caption{A small graph example. 
}\label{fig:R_linkchange}
\end{figure}

To assess the relative sensitivities of Page Rank and reinforcement ranking
to changes in the graph topology, we establish a few useful facts.
First, an important feature of the reinforcement based authority function 
is that it {\em decomposes} over disjoint subgraphs.

\begin{proposition}[Disjoint Independence]\label{thm:union}
For a graph consisting of separate subgraphs,
the $R^\pi$ vector is given by the union of the local $R^\pi$ vectors 
over the disjoint subgraphs. 
(Straightforward consequence of the definition.)
\end{proposition}

As the Web grows, subgraphs are often added that have only limited
connection to the remainder of the web.
In such cases, the $R^\pi$ score remains largely unchanged,
whereas Page Rank is globally affected due to teleportation.
In fact,
merely increasing graph size affects the Page Rank scores for a 
fixed subgraph, since the teleportation vector changes.

Another independence property of reinforcement ranking is that the
$R^\pi$ score for {\em altruistic} subgraphs 
(subgraphs with only outgoing and no incoming links)
is not affected by any external changes to the graph
that do not impact altruism.

\begin{proposition}[Altruistic Independence]\label{thm:altruistic}
The local $R^\pi$ vector for an altruistic subgraph
cannot be affected by external graph changes,
provided no new incoming links to the subgraph are created.
(Immediate consequence of the definition.)
\end{proposition}

Again, Page Rank cannot satisfy altruism independence
due to the global effect introduced by teleportation.

Intuitively, separate websites (i.e.\ separate subgraphs) grow in a nearly
independent manner.
Reinforcement ranking is more stable with respect to independent subgraph 
changes,
since the stationary distribution of Page Rank
must react globally to even local changes.
To illustrate the point, consider the example in 
Figure~\ref{fig:R_linkchange}.
First, suppose the link from $5$ to $6$ is not present;
in which case the graph consists of two disjoint subgraphs.
For reinforcement ranking, any local changes within the subgraphs
(including adding new pages)
cannot affect the authority scores in the other subgraph,
provided no connecting links are introduced between them.
However, the stationary distribution for Page Rank
must be affected even by disjoint updates.
Next, consider the effect of adding a link from $5$ to $6$,
which connects the two subgraphs.
In this case, changes to the right subgraph will still not affect the 
reinforcement scores of the left subgraph if no new
links are introduced from the right to the left, 
whereas Page Rank is affected.
Finally, deleting the link from node $1$ to node $4$ has no influence 
on node $2$ under reinforcement ranking
(only the successors of node $4$ are influenced),
whereas the Page Rank of node $2$ will generally change.

The implication is that the reinforcement based authority score
is more stable to innocuous changes to the Web graph than Page Rank,
which has consequences for both the efficiency of the update algorithms
as well as the quality of their respective authority scores,
as we now demonstrate.
\section{Experimental Results}
\label{sec:exp}

We conducted experiments on real world graphs (Wikipedia and DBLP)
to evaluate two aspects of reinforcement ranking and Page Rank.
First, we compared these methods on the \emph{updating problem}: 
how quickly can the score function be updated
given changes to the underlying graph?
Second, we investigated the overall quality of the score functions produced.

\begin{figure}[t]
\centering
\includegraphics[width=3.2in]{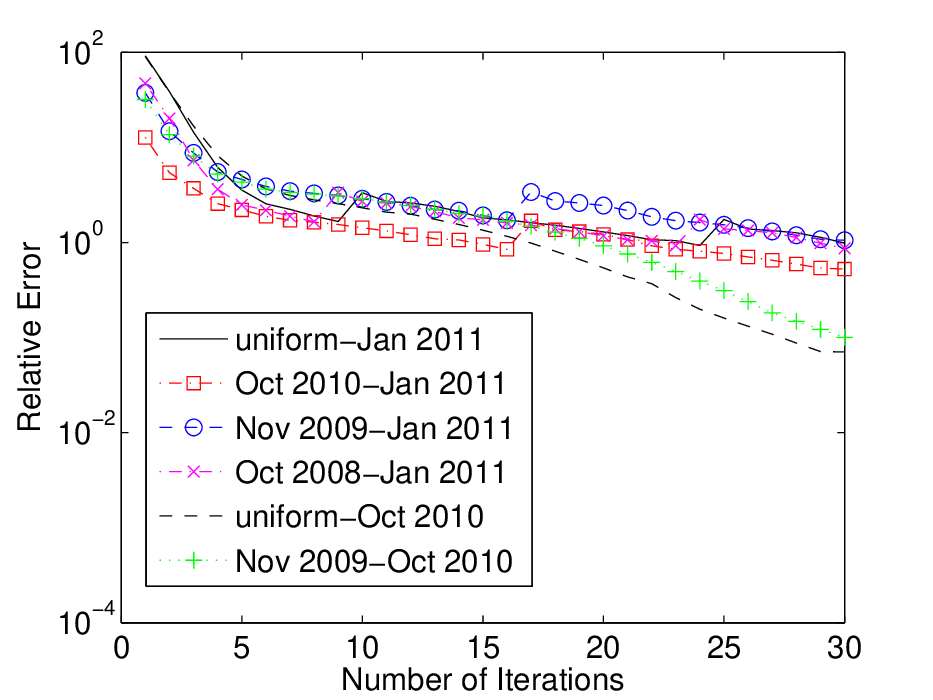}
\caption{
Convergence rate of Page Rank. 
}
\label{fig:errPR}
\end{figure}

\noindent
{\bf Sensitivity and the Updating Problem.}
Intuitively, the speed with which 
an iterative method can update its scores for a 
modified graph is 
related to the sensitivity of its score function.
If the score is not significantly affected by
the graph update, then initializing the procedure from the previous scores
reduces the number of iterations needed to converge.
Conversely, if the new score is significantly different than
its predecessor, one expects that many more iterations will be
required to converge.
Indeed, we find that this is the case:
Page Rank demonstrates far more score sensitivity to graph modification, 
and consequently it is significantly outperformed by reinforcement
ranking in the updating problem.

\begin{figure}[t]
\centering
\includegraphics[width=3.2in]{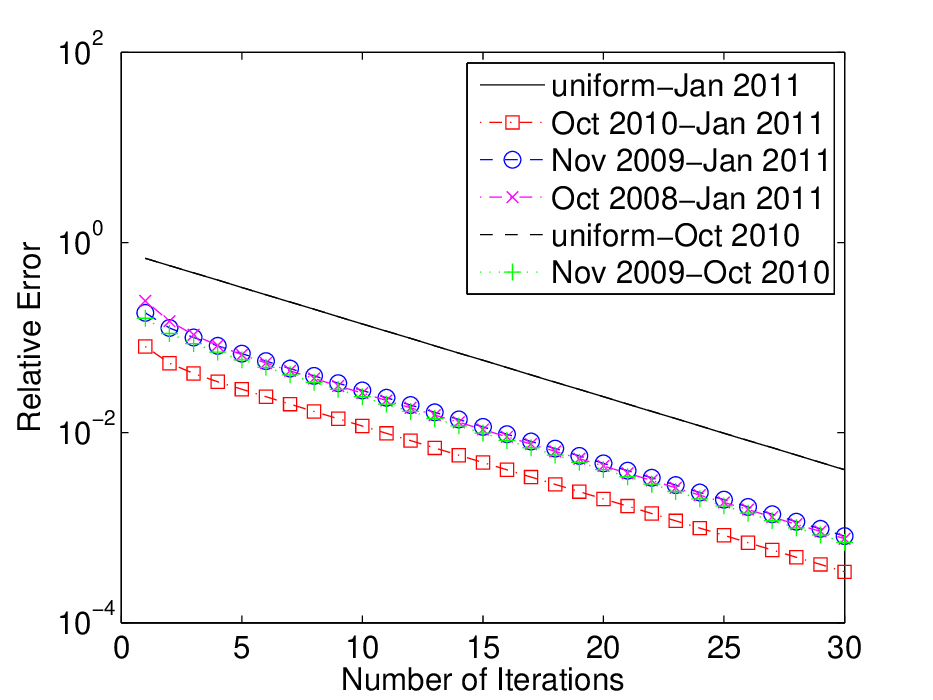}
\caption{
Convergence rate of reinforcement ranking. 
}
\label{fig:errR}
\end{figure}

To investigate this issue, we ran experiments on a set of real world
graphs extracted from Wikipedia dumps taken at different times.
In particular, we used graphs extracted from 
dumps on Oct-2008, Nov-2009, Oct-2010 and Jan-2011.
These are large and densely connected graphs;
for example, the Jan-2011 graph contains $6,832,616$ articles and 
$144,231,297$ links.
For both methods, we used a unform random surfer policy,
and a discount/dampening factor of $0.85$.
For Page Rank, we set the teleportation vector to uniform,
and for reinforcement ranking we used uniform rewards.
To evaluate a given method's 
ability to cope with graph updates, we measured its rate of convergence
to the new solution, as well as the relative advantage of initializing
from the previous solution verus initializing uniformly.
In particular, the plots show the results for the 
(initialization, target) pairs:
\begin{tabular}{rrrll}
\hline
color & initializer & target  & $\Delta$nodes\% & $\Delta$links\% \\
\hline
red     & Oct-2010 & Jan-2011 & +12, -4  & +17, -8\\
green   & Nov-2009 & Oct-2010 & +18, -5  & +39, -20\\
blue    & Nov-2009 & Jan-2011 & +19, -5  & +46, -24\\
magenta & Oct-2008 & Jan-2011 & +49, -18 & +65, -41\\
\hline
\end{tabular}
Here, + indicates the percentage of new nodes/links added,
and - denotes the percentage nodes/links deleted 
between the intial and target graphs.
Figures \ref{fig:errPR} and \ref{fig:errR} compare the 
relative rate of convergence of
Page Rank versus reinforcement ranking. 
%
Note that, given its sensitivity, Page Rank is not able to exploit a previous
solution to significantly improve the time taken to converge to a new
solution for an updated graph: uniform initialization performs as well.
This confirms Google's report that historical update based power iteration 
does not improve the accuracy for Page Rank \cite{amy06}.
By contrast, reinforcement ranking exhibits far less sensitivity
and therefore demonstrates significantly faster convergence when
initialized from a previous graph's score function.
%
Practically this means that, 
initialized with a historical update from three months prior, 
the reinforcement score can be computed about $10$ times more accurately than 
with a uniform initialization.

\noindent
{\bf Ranking Quality.}
To assess the ranking quality of the two methods
we performed an experiment on the DBLP graph \cite{jiecitation}, 
which consists of $1,572,278$ nodes and $2,083,947$ links.
We chose this network because citation links are usually reliable,
reducing the effects of spam and low quality links.
For this experiment, we used the same parameters as before,
except that
for reinforcement ranking 
we used a history depth of 3.

To illustrate the ranking quality achieved by Page Rank
and reinforcement ranking, we show the
highest ranked papers according to each method in Tables \ref{tab:pr_dblp}
and \ref{tab:R3_dblp} respectively.
We used the latest number of citation data retrieved from Google Scholar on March 24, 2013 as the ground truth for paper quality. 
Note that this oracle considers future citations that are received four years later than the time of the link graph was extracted. 
In addition, {Google Scholar} considers much more citation sources than DBLP. 
Although the results exhibit some noise, it is 
clear that the Page Rank scores in Table \ref{tab:pr_dblp} are 
generally inferior:
observe the prevalence of ``outlier'' papers ({\em italicized}) 
that have very few citations. 
By contrast, the reinforcement based ranking in Table \ref{tab:R3_dblp}
completely avoids papers with low citation counts.
Due to the relative purity of the links in this graph, 
it is reasonable to expect a shallow history depth of 3 should be sufficient
to safely identify influential papers in the reinforcement approach.
On the other hand, Page Rank
which considers long term random walks, 
appears to be 
derailed by noise in the graph
and produces more erratic results.


\begin{table}
\begin{scriptsize}
\caption{Top papers according to Page Rank.}
\label{tab:pr_dblp}
\centering
\begin{tabular}{ |l|l|l|}
\hline
Rank & Paper Title & \#Cites\\
\hline
1& {\em A Unified Approach to Functional Dependencies and Relations}&51\\ 
2& {\em On the Semantics of the Relational Data Model}&167\\
3& Database Abstractions: Aggregation and Generalization& 1518\\ 
4& Smalltalk-80: The Language and Its Implementation& 5496\\
5& {\em A Characterization of Ten Hidden-Surface Algorithms}& 847 \\
6& {\em An algorithm for hidden line elimination}& 73 \\
7& Introduction to Modern Information Retrieval& 9056 \\
8& C4& 20913\\
9& Introduction to Algorithms& 30715\\
10& Compilers: Princiles, Techniques, and Tools& 11598 \\
11& Congestion avoidance and control&6078\\
12& A Stochastic Parts Program and Noun Phrase Parser for ...& 1314\\
13& Illumination for Computer Generated Pictures& 2504\\
14& Graph-Based Algorithms for Boolean Function Manipulation& 8252\\
15& Programming semantics for multiprogrammed computations& 777\\
16& Time, Clocks, and the Ordering of Events in a Distributed ...& 7720\\
17& {\em Reentrant Polygon Clipping}& 373\\
18& Computational Geometry - An Introduction& 8558\\
19& A Computing Procedure for Quantification Theory& 2579\\
20& A Machine-Oriented Logic Based on the Resolution Principle& 4077\\
21& Beyond the Chalkboard: Computer Support for Collaboration & 1079\\
22& {\em A Stochastic Approach to Parsing}& 42\\
23& {\em Report on the algorithmic language ALGOL 60}& 646\\
\hline
\end{tabular}
\end{scriptsize}
\end{table}

\section{Discussion}
\label{sec:rel}

A key challenge faced by Page Rank is coping with dangling pages.
Although some dangling pages genuinely do not have any outlinks, 
many are left ``dangling'' simply because crawls are incomplete.
In practice, the number of dangling pages can even dominate 
the number of non-dangling pages \cite{pr_frontier}.  
Page et al.\ (\citeyear{Page98}) 
first removed dangling pages (and the links to them) 
before computing the Page Rank for the remaining graph, 
re-introducing dangling pages afterward.
Such a process, however, does not compute the Page Rank on the original graph.
Moreover, removing dangling pages produces more dangling pages. 
In general, many approaches have been proposed to solve this problem,
but it does not appear to be definitively settled for Page Rank;
see, e.g., 
\cite{pr_frontier,PR_survey}.
This is not a challenge for reinforcement ranking.

Recently, versions of Page Rank have been formulated using linear system 
theory (e.g., see \cite{pr_lin,PR_survey,amy06}).
However, 
the justification for these formulations inevitably returns to
random walks, teleportation, and the resulting stationary distributions. 
As we have observed, such foundations tend to lead to globally sensitive 
ranking methods.
Our work explains and justifies a linear system formulation in a different way.
We generalize the teleportation vector to rewards that evaluate the 
intrinsic importance of individual pages.
Moreover, we have related the linear systems formulation to
work in dynamic programming and reinforcement learning,
via an accumulative rewards-based score function.
It has previously been observed that using a $c$ near $1$ 
in this linear formulation still ``often'' converges, but the reason has not been well 
understood \cite{pr_lin,PR_inout}. 
However, we  have shown that the authority function can be well defined 
and
guaranteed to converge
for any discount factor in $(0,1)$ and any well-defined surfing policy, 
without using teleportation. 

There have been many attempts to formulate
teleportation for more sophisticated ranking, 
such as personalization \cite{Page98,jeh03}, 
query-dependent \cite{qdpr,davood}, 
context-sensitive \cite{taher02,Taherthesis}, 
and battling-link-spam ranking \cite{trustrank}.
For example, the personalized Page Rank surfer teleports 
to the bookmarks of a user. 
However, these practice still rely on the the stationary distribution formulation for convergence. 
In fact, all these can be even more naturally expressed
in a reinforcement ranking framework, and thus convergence guaranteed. 
For example, the preferences of different users can be modeled by
different reward functions over pages, influenced by bookmarks.
(Such reward functions can even be learned via inverse reinforcement
learning, allowing convenient generalization across a large portion
of the graph.)
We can also explain why the pages linked by the bookmarked pages also 
receive a high ranking, 
a fact first observed by \cite{Page98}. 
In particular, the nonzero rewards received by a user on their bookmarked pages 
are also the historical rewards of the successor pages of the bookmarked pages,
hence
the successor pages are also 
rewarded.

\begin{table}
\caption{Top papers according to $R_3$ (3-step history). }\label{tab:R3_dblp}
\label{tab:R3_dblp}
\centering
\begin{scriptsize}
\begin{tabular}{ |l|l|l|}
\hline
Rank & Paper Title & \#Cites\\
\hline
1& C4& 20913\\
 2& Introduction to Algorithms& 30715\\
 3& Introduction to Modern Information Retrieval& 9056\\
 4& Smalltalk-80: The Language and Its Implementation& 5496\\
 5& Compilers: Princiles, Techniques, and Tools& 11598\\
 6& Graph-Based Algorithms for Boolean Function Manipulation& 8252\\
 7& Computational Geometry - An Introduction& 8558\\
 8& Congestion avoidance and control& 6078\\
 9& Time, Clocks, and Ordering of Events in Distributed Sys...& 7720\\
 10& Induction of Decision Trees& 11561\\
 11& Mining Association Rules between Sets ...& 12342\\
 12& A Performance Comparison of Multi-Hop Wireless ...&  4936\\
 13& Fast Algorithms for Mining Association Rules ...& 13827\\
 14& Highly Dynamic Destination-Sequenced ... Routing ...& 6731\\
 15& A Stochastic Parts Program and Noun Phrase ...& 1314\\
 16& Support-Vector Networks& 10523\\
 17& A Machine-Oriented Logic Based on Resolution Principle& 4077\\
 18& A Theory for Multiresolution Signal Decomposition...&  15897\\
 19& An information-maximization approach to blind separation ...& 5871\\
 20& The Anatomy of a Large-Scale Hypertextual Web Search ...& 10122\\
 21& The Complexity of Theorem-Proving Procedures& 4876\\
 22& Combinatorial Optimization: Algorithms and Complexity& 7050\\
 23& A Computing Procedure for Quantification Theory& 2579\\
\hline
\end{tabular}
\end{scriptsize}
\end{table}

\section{Conclusion}
\label{sec:con}
Formuating and viewing Page Rank as the stationary distribution of random walks has been long recognized and practiced. 
However, to gurantee the existence, stochasticity, ergodicity, and irreducibility of the underlying Markov chain has to be ensured. 
This is tricky for the case of Web, where there are many dangling pages, sinks, and pages without any incoming links. 
These problems are important to the theory and practice of Page Rank, for which there are many solutions and discussions. 

We proposed an authority function based on historical rewards. 
We used rewards to capture the intrinsic importance of pages, without the need of teleportating and constructing well behaved Markov chains. 
We related the authority function to the value function in dynamic 
programming and reinforcement learning, 
and showed that the authority function satisfies a reverse Bellman equation. 
Thus, at a high level, our work establishes a theoretical foundation for 
the recent linear system formulation of Page Rank.
We proved that our authority function is well defined for any discount 
factor in $ (0,1)$ and any surfing policy,
by referring not to the stationary distribution theory but to the contration mapping technique.
Given that random walk models, a generalization of Page Rank, 
have been used in various contexts, 
we believe our work will contribute to the fields of information 
retrieval and social networks.

\end{document}